\documentclass[letterpaper]{article} 
\usepackage{aaai24}  
\usepackage{times}  
\usepackage{helvet}  
\usepackage{courier}  
\usepackage[hyphens]{url}  
\usepackage{graphicx} 
\usepackage{natbib}  
\usepackage{caption} 
\usepackage{algorithm}
\usepackage{algorithmic}
\usepackage{footnote}
\usepackage[table,xcdraw]{xcolor}
\usepackage{appendix}

\usepackage{newfloat}
\usepackage{listings}
\DeclareCaptionStyle{ruled}{labelfont=normalfont,labelsep=colon,strut=off} 
\floatstyle{ruled}
\newfloat{listing}{tb}{lst}{}
\floatname{listing}{Listing}
\usepackage{bera}
\usepackage{listings}
\usepackage{xcolor}

\definecolor{eclipseStrings}{RGB}{42,0.0,255}
\definecolor{eclipseKeywords}{RGB}{127,0,85}
\colorlet{numb}{magenta!60!black}

\lstdefinelanguage{json}{
    basicstyle=\normalfont\ttfamily,
    commentstyle=\color{eclipseStrings}, 
    stringstyle=\color{eclipseKeywords}, 
    numbers=left,
    numberstyle=\scriptsize,
    stepnumber=1,
    numbersep=4pt,
    showstringspaces=false,
    breaklines=true,
    string=[s]{"}{"},
    comment=[l]{:\ "},
    morecomment=[l]{:"},
    literate=
        *{0}{{{\color{numb}0}}}{1}
         {1}{{{\color{numb}1}}}{1}
         {2}{{{\color{numb}2}}}{1}
         {3}{{{\color{numb}3}}}{1}
         {4}{{{\color{numb}4}}}{1}
         {5}{{{\color{numb}5}}}{1}
         {6}{{{\color{numb}6}}}{1}
         {7}{{{\color{numb}7}}}{1}
         {8}{{{\color{numb}8}}}{1}
         {9}{{{\color{numb}9}}}{1}
}

\title{A Standardized Machine-readable Dataset Documentation Format \\ for Responsible AI}
\author{
    Nitisha~Jain\textsuperscript{\rm 1*},
    Mubashara~Akhtar\textsuperscript{\rm 1*},
    Joan~Giner-Miguelez\textsuperscript{\rm 2*}, 
    Rajat~Shinde\textsuperscript{\rm 3*},
    Joaquin~Vanschoren\textsuperscript{\rm 6},
    Steffen~Vogler\textsuperscript{\rm 7},
    Sujata~Goswami\textsuperscript{\rm 11},
    Yuhan~Rao\textsuperscript{\rm 13},
    Tim~Santos\textsuperscript{\rm 14},
    Luis~Oala\textsuperscript{\rm 8},
    Michalis~Karamousadakis\textsuperscript{\rm 12},
    Manil~Maskey\textsuperscript{\rm 9},
    Pierre~Marcenac\textsuperscript{\rm 4},
    Costanza~Conforti\textsuperscript{\rm 4},
    Michael~Kuchnik\textsuperscript{\rm 10},\\
    Lora~Aroyo\textsuperscript{\rm 4*},
    Omar~Benjelloun\textsuperscript{\rm 4*},
    Elena~Simperl\textsuperscript{\rm 1, \rm 5*}
}
\affiliations{
    \textsuperscript{*}Core contributors\\
    \textsuperscript{\rm 1}King’s College London,
    \textsuperscript{\rm 2}Universitat Oberta de Catalunya
    \textsuperscript{\rm 3}NASA IMPACT \& UAH,
    \textsuperscript{\rm 4}Google,
    \textsuperscript{\rm 5}Open Data Institute,
    \textsuperscript{\rm 6}TUE \& OpenML,
    \textsuperscript{\rm 7}Bayer,
    \textsuperscript{\rm 8}Dotphoton,
    \textsuperscript{\rm 9}NASA,
    \textsuperscript{\rm 10}Meta,
    \textsuperscript{\rm 11}Oak Ridge National  Laboratory,
    \textsuperscript{\rm 12}Plaixus Ltd,
    \textsuperscript{\rm 13}North Carolina State University,
    \textsuperscript{\rm 14}Graphcore

,

}

\usepackage{bibentry}

\begin{document}

\maketitle

\begin{abstract}
Data is critical to advancing AI technologies, yet its quality and documentation remain significant challenges, leading to adverse downstream effects (e.g., potential biases) in AI applications. This paper addresses these issues by introducing Croissant-RAI, a machine-readable metadata format designed to enhance the discoverability, interoperability, and trustworthiness of AI datasets. Croissant-RAI extends the Croissant metadata format and builds upon existing responsible AI (RAI) documentation frameworks, offering a standardized set of attributes and practices to facilitate community-wide adoption. 
%
%
Leveraging established web-publishing practices, such as Schema.org, Croissant-RAI enables dataset users to easily find and utilize RAI metadata regardless of the platform on which the datasets are published. Furthermore, it is seamlessly integrated into major data search engines, repositories, and machine learning frameworks, streamlining the reading and writing of responsible AI metadata within practitioners' existing workflows. 
Croissant-RAI was developed through a community-led effort. It has been designed to be adaptable to evolving documentation requirements and is supported by a Python library and a visual editor.
\end{abstract}

\section{Introduction}

Data is a crucial element in emergent AI technologies and plays a central role in training and evaluating models. Recent research has emphasized how data can unexpectedly lead to harmful outcomes in AI applications, highlighting the real-world consequences of data-related issues. For instance, recent studies have revealed gender-biased AI classifiers in computer-aided diagnosis as a direct result of imbalances in training data~\cite{larrazabal2020gender}, and issues during data collection that can cause AI applications, which are designed to detect pneumonia, to fail when deployed to different hospitals~\cite{zech2018variable}.

Consequently, the community focused on responsible AI (RAI) has identified data work as critical to the development of trustworthy AI systems~\citep{Smuha2019,2024dmlr}. 
Seminal works, such as Datasheets for Datasets~\cite{gebru2021datasheets} and Data Statements~\citep{bender2018data} have emphasized the importance of dataset documentation to assess and increase the trustworthiness of AI systems.
Data needs to be documented, analyzed, and enriched (e.g., against biases), and maintained. 

While the research community has made significant progress in publishing and sharing datasets, most datasets are still one-time, expensive efforts~\cite{10.1145/3411764.3445518}. Formats for documenting AI data can be a rich source of information, but it is becoming increasingly clear that they need to evolve \cite{yang2024navigating}. 
Current proposals have overlapping formats, require data documentation written in natural language, lack a standard structure and are not integrated with widely used tools and ML frameworks, making them difficult for machines to read and process. 

\begin{savenotes}
\renewcommand{\arraystretch}{1.0}
\begin{table*}[th]
    \centering
    \caption{Dataset documentation toolkits used as base for the vocabulary design \\}
    \label{tab:data_documentation_toolkits}
    
    \begin{tabular}{p{6cm}p{9.8cm}}
       \hline  
        \textbf{Toolkit} & \textbf{Description} \\ \hline   
        HuggingFace Dataset Cards\footnote{\url{https://huggingface.co/docs/hub/en/datasets-cards}} & HuggingFace encourage data authors to publish natural text reports following a similar structure as Datasheets for Datasets~\citep{gebru2021datasheets}. \\    
        \hline
        Kaggle metadata\footnote{\url{https://github.com/Kaggle/kaggle-api/wiki/Dataset-Metadata}} & Kaggle platform allows users to upload composition and contextual metadata along datasets. \\ 
        \hline
        Data Nutrition Labels~\cite{holland2018dataset} & The Data Nutrition Project aims to create a standard label for interrogating datasets, inspired by Datasheets for Datasets. \\ 
        \hline
        Data Cards \cite{pushkarna2022data} & Data Cards are structured summaries of critical details about various elements of ML datasets that participants need throughout a project's lifescycle for ethical AI development. \\ 
        \hline
        Croissant \cite{2024croissant} & Croissant is a metadata format for machine learning datasets that combines metadata, resource file descriptions, data structure, and default ML semantics into a single file. \\ 
        \hline
        Crowdworksheets~\cite{diaz2022crowdworksheets} & A framework to facilitate transparent documentation of key decisions points at various stages of the annotation pipeline focusing on crowdworkers.\\ 
        \hline
        Fairness Datasets Ontology\footnote{\url{https://fairnessdatasets.dei.unipd.it/schema/}} & The Fairness Datasets Ontology clearly outlines the relationships among various fields in the data briefs and explicitly details the connections to external vocabularies. \\ 
        \hline
        DescribeML~\cite{giner2023domain} & DescribeML is a domain-specific language (DSL) to precisely describe machine learning datasets in terms of their structure, provenance, and social concerns. \\ 
        \hline
      
    \end{tabular}
\end{table*}
\end{savenotes}

This work addresses these issues by introducing \textbf{Croissant-RAI}, a machine-readable format for capturing and publishing RAI-related data documentation.
It consists of a set of attributes\footnote{In this manuscript, the terms \textit{attributes} and \textit{properties} are used interchangeably. Typically, in Computer Science, attributes refer to additional information, whereas properties describe the characteristics of the object. However, in English, both have similar connotations in terms of defining characteristics of a real-world concept.} organized around RAI use cases and related documentation to promote community-wide adoption. 
It builds on and complements existing RAI dataset documentation proposals, making publishing, discovering, and reusing existing RAI documentation easier. Croissant-RAI is an extension of the Croissant metadata format~\citep{2024croissant}. Croissant is built on established web-publishing practices, such as Schema.org~\cite{guha2016schema}, and designed to improve ML datasets' discoverability, portability, reproducibility, and interoperability. It makes datasets discoverable by data search engines and enables loading them directly into ML frameworks and tools.

Croissant-RAI was developed through a multi-step vocabulary engineering process based on recent RAI data documentation frameworks and inspired by practical use cases. These use cases cover data design and collection processes, human and machine data labeling, data participatory processes, AI safety and fairness assessments, and data regulatory compliance with current AI regulations. Croissant-RAI is governed by a community-led effort and intends to be a dynamic asset for addressing new documentation requirements. Finally, Croissant-RAI is accompanied by a Python library and a visual web editor to assist users during its usage.




The rest of the paper is organized as follows: First, we discuss background related to machine-readable metadata, the Croissant format, and state-of-the-art RAI documentation frameworks. Then, we present the vocabulary engineering process and the use cases central to Croissant-RAI. We further introduce the Croissant-RAI vocabulary together with dataset examples demonstrating its use. Finally, we present the tool support around Croissant-RAI and discuss the implications of this work.  


%

\section{Background}
\label{back}

\begin{figure*}[h!] 
\centering
\includegraphics[width=2\columnwidth]{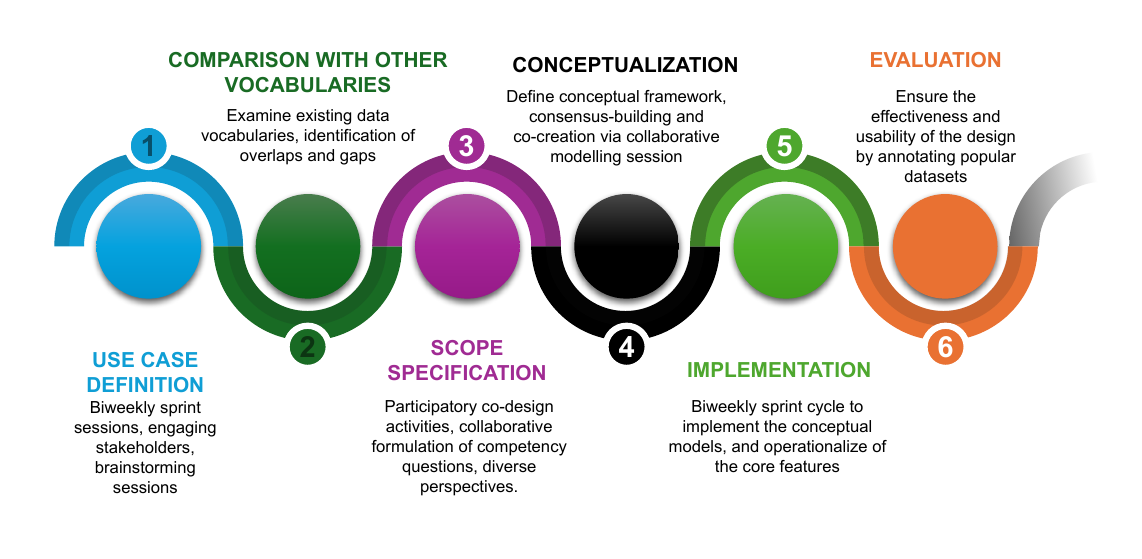}
 \caption{The Multi-step Vocabulary Engineering Process for Croissant-RAI}
 \label{fig:desing_steps}
\end{figure*}

This section discusses related background on machine-readable metadata, the Croissant metadata format, and responsible AI.

\subsection{Machine-readable metadata}

As more data becomes available online, the scientific community has long focused on how to make this data discoverable and easy to use. 
To accomplish this, the standard approach is to supplement the published data with the relevant machine-readable metadata. This metadata may include information about the data's repository, authors, or specific tags, expressed in a format that sticks to consistent structures to facilitate machine computation. For example, the Data Catalog Vocabulary (DCAT)~\cite{albertoni2023w3c} is a metadata format that enables interoperability between web-based data catalogs, allowing users to automatically aggregate, classify, and filter data. Institutions like the European Union use this standard to develop open data portals that automatically integrate data from multiple sources. These portals ensure data compliance with the FAIR principles~\cite{wilkinson2016fair}, demonstrating the benefits and practical applications of machine-readable metadata.

Other metadata standards aim to improve the content discoverability across the web. For example, Schema.org \cite{guha2016schema} is a de facto metadata standard designed to help search engines discover structured content on the Web. Search engines can find and index published content described with Schema.org, increasing understandability of content. Furthermore, Schema.org is a versatile format that can describe a variety of content. For instance, content described with \texttt{Schema.org/Dataset}\footnote{\url{http://schema.org/Dataset}} extension is understood and indexed as a dataset by search engines like Google Dataset Search\footnote{\url{https://datasetsearch.research.google.com/}}.

\subsection{Croissant: A metadata format for ML-ready datasets}

Machine-readable metadata formats are gaining popularity in the ML community. Croissant~\cite{2024croissant} is a new metadata format for ML-ready datasets, developed through a community-led effort involving both industry and academia. Croissant is designed to improve discoverability, portability and reproducibility of ML datasets. Datasets described with Croissant and published online can be easily discovered by current data search engines, regardless of their publication location. Additionally, Croissant prepares datasets for ML use cases by allowing them to be directly loaded into frameworks like PyTorch, TensorFlow, or JAX, streamlining data loading across frameworks.

Croissant extends Schema.org with additional layers that comprehensively describe datasets' attributes, resources, structure, and semantics to facilitate their use and sharing within the ML community. The Croissant format does not change dataset content representation (e.g., image or text file formats) but provides a standard way to describe and organize data, facilitating adoption by a broad community.

\subsection{Responsible AI data documentation} 


Over the past years, data work is increasingly seen as critical to the development of trustworthy AI systems, considering aspects related to accountability, fairness, transparency, data privacy and governance, robustness and safety~\citep{Smuha2019}.
While the field has seen progress in data publishing and sharing, more often than not data work is considered as a one-off effort.
To address its potential negative downstream effects, data needs to be maintained, documented, analysed and enriched (e.g. against biases)~\citep{10.1145/3411764.3445518}. 

%
Previous works have emphasized the importance of dataset documenting to develop fairer and more trustworthy AI applications. Documentation efforts such as Datasheets for Datasets~\citep{gebru2021datasheets}, Data Nutrition Labels~\citep{holland2018dataset}, and Data Statements~\cite{bender2018data} have encourage data publishers to document data dimensions that may affect dataset use or the quality of resulting ML models.
Seminal works on data documentation~\citep{bender2018data, gebru2021datasheets} have inspired the development of various tools for data documentation (see overview in Table~\ref{tab:data_documentation_toolkits}). For example, Huggingface Dataset Cards and Data Nutrition Labels have been inspired by Datasheets for Datasets which documents a dataset's motivation, creation, composition, intended uses, distribution, maintenance, and further information~\citep{gebru2021datasheets}.

However, these approaches and tools require data documentation primarily written in natural language, lack a standard structure and miss integration with widely used tools and ML frameworks. This makes data documentation difficult for machines to read and process. Moreover, the proposed dimensions do not support discoverability by search engines and are challenging to integrate in frameworks and tools used across the entire life-cycle of an AI system. 
This work addresses these issues by proposing a machine-readable framework for capturing RAI data documentation practices as an extension of the Croissant format, making RAI-related data documentation more discoverable, maintainable and interoperable.

\section{Croissant-RAI Vocabulary Engineering}
\label{method}

This section presents the details of the vocabulary design and engineering process for Croissant-RAI format. The entire process has been led by community projects involving experts from various backgrounds, referred to as stakeholders below. The different aspects of Croissant-RAI were meticulously defined using a comprehensive vocabulary engineering process which was focused on iterative and participatory co-design activities. In this section, we explain the process in detail, including the RAI documentation frameworks that served as the foundation for developing the vocabulary. The subsequent section will discuss the use cases developed to guide the design of Croissant-RAI ensuring its usefulness and relevance to machine learning practitioners.

Figure \ref{fig:desing_steps} depicts the different steps of the vocabulary engineering process detailed as follows: 


\begin{itemize}
\item \textbf{Use Case Definition.} The process began with biweekly sprint sessions, engaging stakeholders from various domains and levels of expertise. Through several brainstorming sessions, stakeholders identified and prioritized the attributes for the Croissant-RAI format. This collaborative approach ensured that the identified use cases were comprehensive, relevant, and aligned with the overarching goals of promoting data standardization and responsible AI practices. \looseness=-1

\item \textbf{Comparison with existing dataset documentation vocabularies.} In this phase, existing dataset documentation vocabularies were thoroughly examined to identify overlaps and gaps in relation to the Croissant vocabulary, i.e. the attributes.
An overview of these vocabularies is presented in Table~\ref{tab:data_documentation_toolkits}. Stakeholders were engaged in comparative analyses and discussions to assess the suitability of different vocabularies for integration with Croissant-RAI.

\item \textbf{Scope Definition.} Through participatory co-design activities, stakeholders formulated competency questions to that defined the scope and requirements of the Croissant-RAI vocabulary. These questions reflected diverse perspectives and priorities, acting as guiding principles. They helped identify the key features and functionalities necessary to meet the objectives of the Croissant-RAI vocabulary.

\item \textbf{Conceptualization of RAI extension and Implementation.} Building on insights from co-design activities discuss above, the conceptual framework for the Croissant-RAI vocabulary was defined on top of Croissant. Through collaborative modeling sessions, concept mapping exercises, and scenario-based discussions, we ensured that the Croissant-RAI vocabulary aligned with use cases, stakeholder priorities and did not overlap or contradiction with the Croissant format.
Moreover, in biweekly sprint cycles a conceptual model of the Croissant-RAI vocabulary as developed. This facilitated an integration of Croissant-RAI with Croissant and ensured compatibility with existing functionalities.




\item \textbf{Evaluation through example annotations for each use case.} To ensure the effectiveness and usability of the Croissant-RAI metadata format, stakeholders engaged in participatory evaluation activities. This included annotating sample datasets using the implemented features and gathering feedback from users to identify any usability issues or areas for improvement.

\begin{figure*}[h!] 
\centering
\includegraphics[width=2\columnwidth]{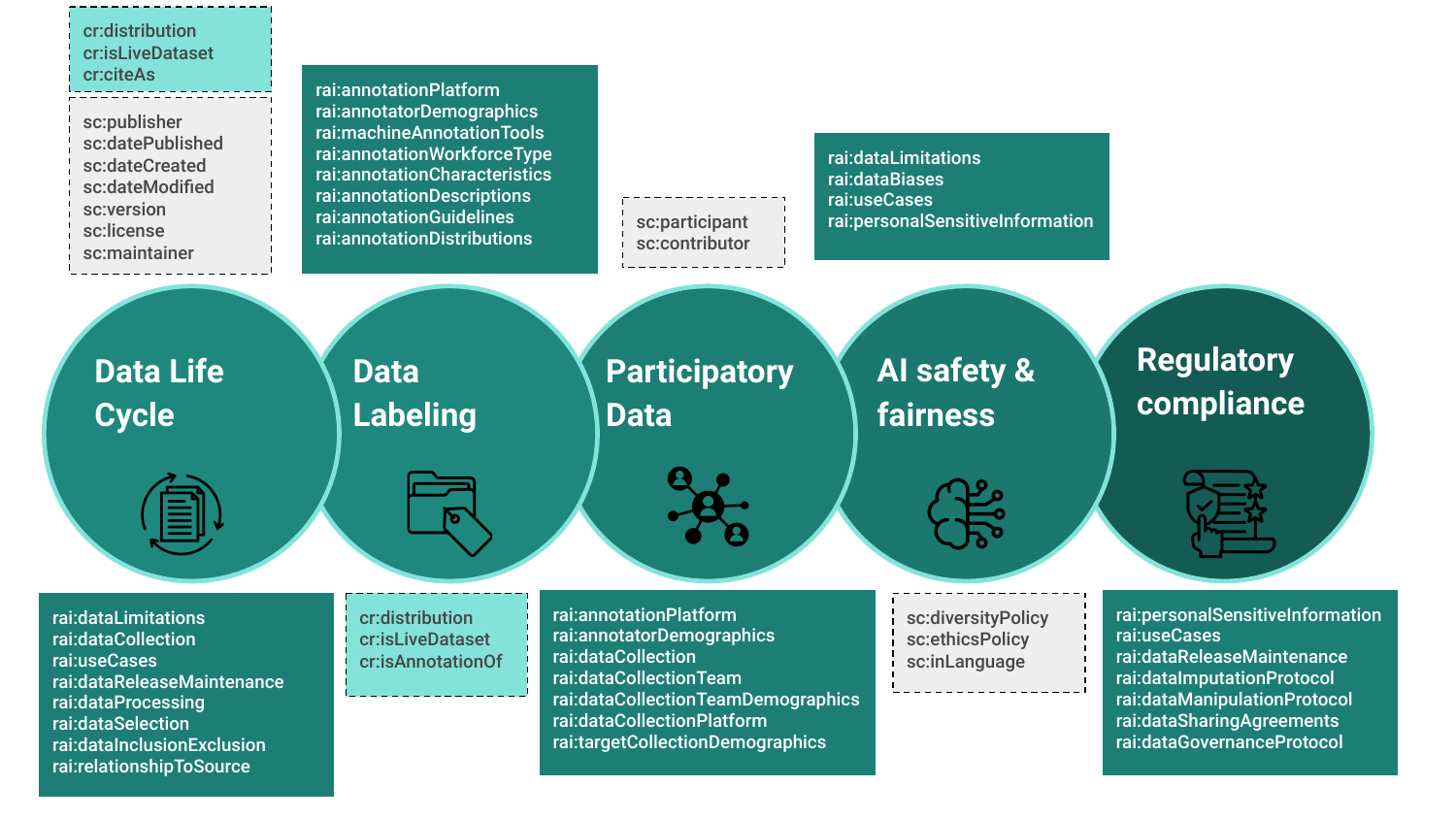}
 \caption{Mapping between the use cases and the properties of the Croissant-RAI vocabulary, Croissant and Schema.org. 
 Dotted boxes include Schema.org (sc:) and Croissant (cr:) properties while the other boxes contain Croissant-RAI properties (rai:).  
 }
  \label{fig:use_cases} 
\end{figure*}

\end{itemize}

Throughout each of these steps, participatory co-design activities played a pivotal role in ensuring that the features of Croissant-RAI were aligned with the needs and perspectives of diverse stakeholders. By fostering collaboration and inclusivity, the vocabulary engineering process facilitated the development of a robust and adaptable framework for promoting standardization and responsible AI practices within the data documentation ecosystem. Note that this engineering process is scalable to other domain-specific extensions that could be developed on top of Croissant in the near future.

\section{Use Cases}
\label{cases}

A set of use cases guided the design of the attributes for the RAI croissant format. During several brainstorming sessions, stakeholders identified and prioritized the five use cases presented in this section. The collaborative approach ensured that the identified use cases were comprehensive, relevant, and aligned with the overall goals of promoting data standardization and responsible AI practices. The presented use cases are intended to be a representative sample, but more may be added in the future.

The initial use cases are centered on documenting the data life cycle, characterizing in-depth data labeling and data participatory processes, and providing critical information for AI safety and fairness assessments and regulatory compliance checks. A more detailed explanation is provided below.

\subsubsection{Use case 1: The data life cycle.}

The dataset life cycle involves stages like motivation, composition, collection, preprocessing, uses, distribution, and maintenance. Documenting RAI-related properties prompts creators to reflect on the process and aids user understanding. Information includes: creator, purpose, creation date, data sources, versioning, composition, processing, intended use, and maintenance. Documentation of  the provenance and lineage of the datasets that are derived from revision, modification or extension of existing datasets is also relevant for this use case.

\subsubsection{Use case 2: Data labeling.  }
Labels plays a central role in dataset for AI and can be obtained through human input such as labels from crowdsourcing platforms, or machine annotations like concept extraction. Details such as the platform used, number of human labels per record, rater demographics, and annotation tool characteristics aid in understanding dataset composition and facilitate efficient sampling. Information on the labeling process enhances comprehension of data creation, sample characteristics, and aids in assessing, replicating, and reproducing the process, thereby boosting data reliability.

\subsubsection{Use case 3: Participatory data.}
Some ML datasets are created using fairly well understood, albeit poorly documented, processes and practices. Others, however, are the result of community or collaborative work that involve a much wider range of entities with limited coordination among them. Examples include citizen science datasets created through participatory sensing; Wikidata, created by 23k editors; or ML datasets using crowdsourcing platforms. Documenting the participatory element of the dataset life cycle can help understand biases and limitations in the dataset, making the process easier to monitor, assess, repeat, replicate, and reproduce.

\subsubsection{Use case 4: AI safety and fairness evaluation.}
Safety and bias information is crucial for understanding potential risks and fairness aspects of data usage, aiming to prevent unintended harmful consequences from models trained on or evaluated with the data. This includes identifying features for known and intended usage (e.g., adversarial datasets for safety evaluation, counterfactual annotations for fairness evaluation) and any usage restrictions. Additionally, accounting for personal and sensitive information within datasets aids in risk mitigation and responsible usage. This information is typically gathered at the item level and aggregated at the dataset level, often presented as a scorecard or nutrition label.

\subsubsection{Use case 5: Regulatory compliance }
Compliance officers and legal teams require data-related information to assess the dataset's fit to privacy and the current regulation laws. For instance, regulations in the field of AI, such as the recently approved European AI Act, state the need to provide documentation about the data used to train specific ML applications\footnote{\url{https://www.euaiact.com/annex/4}, section 2.d}. In regards to this, the Croissant-RAI vocabulary allows users to annotate this information in a structured way, including but not limited to, \textit{sensitive and personal identifiable information}, \textit{data governance} and \textit{security measures and data sharing agreements}. This information can be relevant across various fields, including research, businesses, and the public sector.

\section{The Responsible AI Vocabulary}
\label{vocab}

\begin{figure}[t!] 
\centering
\includegraphics[width=1\columnwidth]{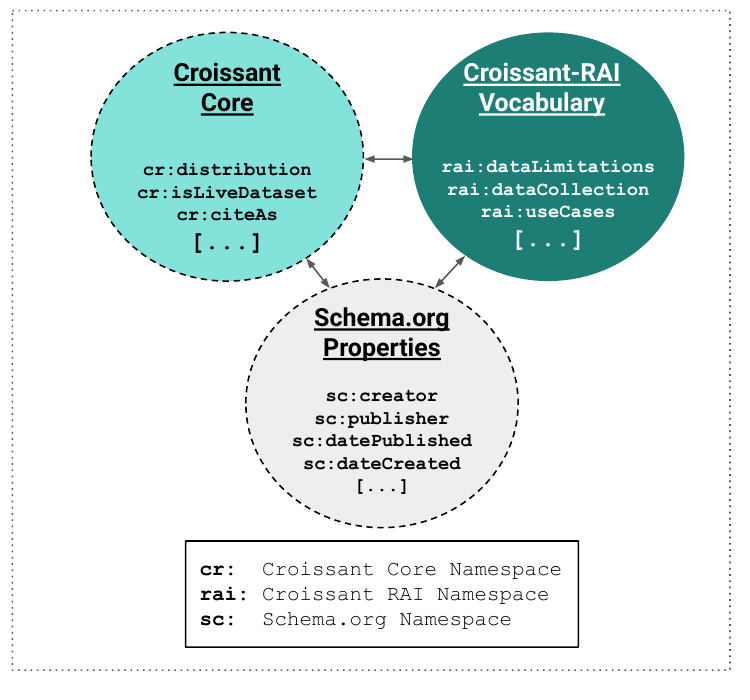}
 \caption{Overview of the Croissant-RAI vocabulary}
 \label{fig:rai_vocab}
\end{figure}

In this section, we present the attributes that comprise the Croissant-RAI vocabulary. The attributes are presented alongside the corresponding use cases discussed earlier. As an extension of Croissant, which is based on the Schema.org/Dataset vocabulary, the Croissant-RAI extension's properties incorporate both core Croissant and Schema.org properties. 
Figure~\ref{fig:rai_vocab} shows the relationship between these vocabularies and a few representative properties of each one.

A concrete mapping of the Croissant-RAI properties to the five use cases identified during the engineering process is shown in Figure~\ref{fig:use_cases}. The figure identifies the different properties related to each use case and their source vocabulary, i.e., whether the property was derived from Croissant core (marked with "cr:") or Schema.org (marked with "sc:") vocabularies or if it is defined as part of the RAI extension (marked with "rai:").\footnote{A detailed description of each property of the vocabulary is provided in the appendix (Table~\ref{tab:descriptions}).} 

In the \emph{Data Life Cycle} scenario, existing information from Schema.org and the Croissant vocabulary covers aspects like creators, publishers, and licenses. However, the Croissant-RAI extension requires additional details about the data collection process, design assumptions including data use cases and identified limitations, and specific data preprocessing steps.


Conversely, in the \emph{Data Labeling} use case, the Croissant-RAI vocabulary offers attributes to ensure label quality. Essential details include the annotation process and annotator information. Documenting instructions given to annotators, annotation devices used, and annotator demographics become crucial for assessing label quality.


For the \emph{Participatory Data} use case, it is crucial to document information about the data collectors, including their demographics and the platforms they use, such as mobile applications. Additionally, if the data represents or is collected from individuals, it is important to gather demographic details about the target participants.


In the context of \emph{AI Safety \& Fairness}, the Croissant-RAI suggests documenting data limitations, biases, and sensitive information within the dataset. These attributes encourage data authors to reason about their data with a specific schema in mind, akin to structure of proposals such as Datasheets for Datasets.


Lastly, concerning \emph{Regulatory Compliance} use case, we propose annotating information about handling sensitive data for privacy, data maintenance policies, intended use cases (aligned with the Use cases classification of the AI Act), and various sharing agreements and manipulation protocols applied to the data.

In the following section, we showcase a collection of datasets described using the Croissant-RAI vocabulary, demonstrating its practical application.



%

\begin{lstlisting}[language=json,numbers=none, basicstyle=\scriptsize	, upquote=true,
caption= Excerpt description of the HLS Burn Scar Scenes dataset, float=t, label=geo]
{
  "@type": "schema.org/Dataset",
  "name": "HLS Burn Scar Scenes",
  "dct:conformsTo": "mlcommons.org/croissant/RAI/1.0",
  "rai:dataCollection": "Imagery is collected from V1.4 of Harmonized Landsat and Sentinel-2 (HLS). A full description and access to HLS may be found at https://hls.gsfc.nasa.gov/. The labels were from shapefiles maintained by the Monitoring Trends in Burn Severity (MTBS) group. The masks may be found at: https://mtbs.gov/",
  "rai:dataProcessing": "After co-locating the shapefile and HLS scene, the 512x512 chip was formed by taking a window with the burn scar in the center. Burn scars near the edges of HLS tiles are offset from the center. Images were manually filtered for cloud cover and missing data to provide as clean a scene as possible, and burn scar presence was also manually verified.",
  [...]
}
\end{lstlisting}



\begin{lstlisting}[language=json,numbers=none, basicstyle=\scriptsize	, upquote=true,
caption= Excerpt description of DICES-350 dataset, float=t, label=DICES]
{
  "name": "DICES-350",
  "dct:conformsTo": "mlcommons.org/croissant/RAI/1.0",
  "rai:dataAnnotationProtocol": "The annotation task included the following six sets of questions: Q1: addresses the whole conversation and asks the raters to assess the legibility of the conversation - is it (1) in English, (2) comprehensible, (3) on a topic that the rater is familiar with or (4) none of the above. Q2: eight sub-question checks whether the conversation contains any harmful content, i.e., whether it could directly facilitate serious and immediate harm to individuals, groups ...",
  "rai:dataAnnotationPlatform" : "Crowdworker annotators with task specific UI",
  "rai:dataAnnotationAnalysis": "Initial recruitment of 123 raters for the DICES-350 dataset, after all annotation tasks were completed, a quality assessment was performed on the raters and 19 raters were filtered out due to low quality work ..."
  "rai:annotatorDemographics": "DICES-350 was annotated by a pool of 104 raters. The rater breakdown for this pool is: 57 women and 47 men; 27 gen X+, 28 millennial, and 49 gen z; and 21 Asian, 23 Black/African American, 22 Latine/x, 13 multiracial and 25 white. All raters signed a consent form agreeing for the detailed demographics to be collected for this task."
  [...]
}

\end{lstlisting}

\section{Application examples}
\label{examples}

This section contains three examples of popular and diverse datasets described in Croissant-RAI format. This goal is to demonstrate the applicability of the Croissant-RAI properties to datasets from various fields in order to train and evaluate AI models. We present a geospatial dataset \cite{HLS_Foundation_2023} with relevant RAI details in the data collection and preprocessing, a dataset for evaluating diversity in conversational AI, documenting the great diversity within the annotators profiles, and a community-created text dataset used to train one of the most recent open-source large language models.

\begin{lstlisting}[language=json,numbers=none, basicstyle=\scriptsize	, upquote=true,
caption= Excerpt description of BigScience Roots Corpus dataset, float=t, label=rots]
{
  "name": "BigScience Root Corpus",
  "dct:conformsTo": "mlcommons.org/croissant/RAI/1.0",
  "rai:dataLimitations": [
    0:"Crawled content over-represents pornographic text across languages in the form of spam ads...",
    1:"The preprocessing removes some categories of PII but is still far from exhaustive ...",
    2:"The reliance on medium to large sources of digitized content still over-represents privileged voices and language varieties."
  ],
  "rai:dataBiases": "Dataset includes multiple sub-ratings which specify the type of safety concern, such as type of hate speech and the type of bias or misinformation, for each conversation ...",
  "rai:dataManipulationProtocol":[
    0:"HTML reconstruction: A 20 fairly simple heuristic ontag types to reconstruct the structure of the text extracted from an HTML code",
    1:"Data filtering: Documents were filtered by a set of criteria such as; Too high character repetition, too high ratio of spatial characters, to high ratio of flagged words to avoid pornographical conctect ...",
    2:"Deduplication: Substring deduplication (Lee et al., 2022) based on Suffix Array (Manber and Myers, 1993) as a complementary method that clusters documents sharing a long substring, were applied for documents with more than 6000 characters."
  ]
  [...]
}

\end{lstlisting}

\begin{figure}[h!] 
\centering
\includegraphics[width=1\columnwidth]{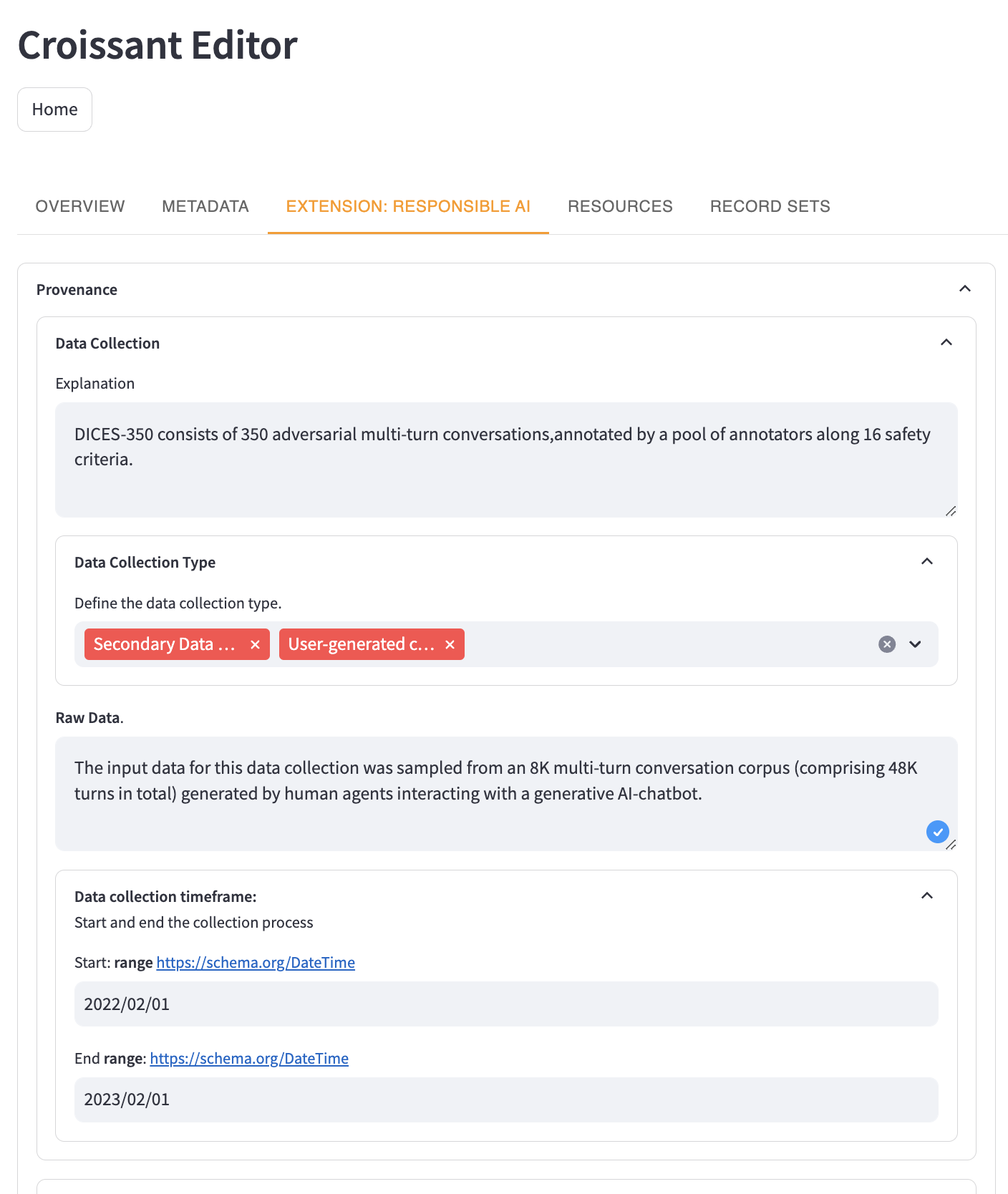}
 \caption{UI of the RAI extension of the Croissant editor}
 \label{fig:editor}
\end{figure}

\subsection{RAI Geospatial AI-ready datasets}

Geospatial AI (also GeoAI) refers to the integration of AI techniques with geospatial data, enabling advanced location-based analysis, mapping, and decision-making. The GeoAI datasets are curated from data captured by various sensors on-board spaceborne, airborne, and ground platforms along with the in-situ sensors. With the ever rising significance of location-based analysis and the increasing size of the generated data, GeoAI plays an important role in solving complex and crucial problems like weather forecasting. However, Geospatial datasets also involve ethical considerations in the acquisition and utilization of geospatial data, addressing potential biases, environmental impact, and privacy concerns.


For instance, location accuracy and spatial properties are vital to the credibility of GeoAI-ready datasets. Changes in local accuracy can have an impact on AI predictions. For example, in AI-based crop yield prediction tasks, ground truth for validating the AI model is typically obtained from agricultural farms. However, due to privacy concerns, these annotations are often approximated, which can result in inaccurate predictions and estimates of AI models. In this regard, RAI properties related to data collection, such as the devices used and details about data preprocessing and manipulation, can increase support and confidence in developed AI models.


In Listing \ref{geo} we can see an excerpt of the HLS Burn Scar Scenes dataset \cite{HLS_Foundation_2023}, an AI-ready geospatial dataset described using the Croissant-RAI format. The listing shows information about the devices used to collected the satellite images, and the preprocesses applied to center the detected burn scars. Other RAI relevant attributes for Geospatial AI dataset such as the sampling strategy or the data validity timeframe (e.g., seasonal agricultural crop yield estimation) can also be annotated using the Croissant-RAI format.

\subsection{Diversity In Conversational AI Evaluation for Safety}

With the increasing rise in development of the LLMs, it is crucial to assess the LLMs for inclusiveness and FAIRness in an operational scenario \cite{vidgen2024introducing}. This also holds true for the conversational AI agents and the chatbots that automate human-computer interaction based on question-answering, data captioning and text retrieval tasks. The DICES (Diversity In Conversational AI Evaluation for Safety) dataset~\cite{aroyo2024dices} addresses the need for nuanced safety evaluation in language modeling, emphasizing diverse perspectives in model training and assessment.

It introduces several key features, including a focus on rater diversity, where differences in opinions are viewed in terms of diversity rather than bias. The dataset ensures balanced demographic representation among raters and assesses safety across five categories, offering detailed evaluations of harm, bias, misinformation, politics, and safety policy violations.


Listing \ref{DICES} contains an excerpt from the Croissant-RAI description of the DICES dataset showing some dataset key features. The excerpt discusses the data annotation protocol, the platform, and an analysis of the annotation process. Furthermore, the property \emph{rai:annotatorDemographics} provides demographic information showing the diversity and inclusion effort done by dataset creators in accommodating language characteristics across gender, race, and age groups. Such metadata-defined information fosters trust in using the data to create conversational AI models capable of generating responses to diverse and FAIR audience groups.

\subsection{BigScience Roots Corpus dataset}

Due to the popularity of language models, there is an increasing interest in large, high-quality datasets, particularly in multilingual contexts. Recent open-source efforts have been focused on creating these kinds of datasets, providing them to train large-scale monolingual and multilingual AI models and encouraging further research in the field. The BigScience workshop\footnote{BigScience workshop homepage: \url{https://bigscience.huggingface.co}}, a year-long international and interdisciplinary effort with the primary goal of investigating and training large language models, has led one of these efforts focusing on ethical considerations, potential harm, and governance issues.

As a result, they generated the BigScience Roots Corpus \cite{laurenccon2022bigscience}, a large-scale dataset collected from multiple sources that helped to train the BigScience Large Open-science Open-access Multilingual (BLOOM) language model, which has 176 billion parameters. The critical points of this dataset are its inherent variate composition as they gathered data from multiple sources, the inherited limitations and biases from these sources, and the set of different preprocessing steps applied to ensure its quality. 

Listing \ref{rots} contains an excerpt from the BigScience Roots Corpus dataset Croissant-RAI description that describes some of the detected data limitations and biases from the sources, as well as some of the preprocessing steps applied to the data. Sharing biases and data limitations helps potential users evaluate the data's suitability for their use cases, and sharing the preprocessing steps used on the data improves the reproducibility of dataset creation by future similar initiatives.

\subsection{Tool support}

To boost and facilitate the community's adoption of the Croissant-RAI format, we have developed an extension of the current Croissant python library and its web UI editor to implement the Croissant-RAI vocabulary. This editor allows users to input RAI attributes in JSON-LD form, thus serving as a tool for data publishers to input dataset-related RAI information.
Figure~\ref{fig:editor} shows the Croissant Editor's RAI tab, which currently defines a few RAI metadata attributes for the dataset publishers. Once the required information is populated, the data publishers can export the final Croissant metadata as the output, containing the supported RAI attributes that have been presented in this work.

\section{Conclusions}
\label{conclusions}


In this work, we present Croissant-RAI, a machine-readable format for capturing and publishing RAI documentation meta-data for AI datasets. Our proposal has been built on top of previous data documentation approaches, making them easy to share, discover, and reuse. The design of the vocabulary has been derived through relevant RAI use cases and illustrated in real examples from important domains and datasets. Finally, we expand the current Croissant toolkit, implementing the Croissant-RAI extension to aid the community in its adoption.


In future work, we aim to track the community's uptake of the Croissant-RAI vocabulary, offering valuable insights into its real-world application. Moreover, we recognize the socio-technical dimensions of RAI practices and underscore the importance of collaborating with both public and private partners. Through active engagement with regulators and corporations, we can not only promote the widespread adoption of RAI methodologies but also ensure their responsible implementation. We maintain that institutions should contemplate integrating RAI benchmarks, such as the one outlined in this paper, into their frameworks to foster responsible AI practices.
%




\bibliography{aaai24}


\appendix
\section{Appendix}

\begin{appendices}

Table~\ref{tab:descriptions} gives an overview of all attributes of Croissant-RAI. 



\renewcommand{\arraystretch}{1.3}
\begin{table*}[h]
    \centering
    \tiny
    \caption{Croissant-RAI Property Descriptions\\}
    \label{tab:descriptions}
    \begin{tabular}{p{3cm}p{1.2cm}p{2cm}p{1cm}p{8.5cm}}

       \hline
  
\textbf{Property} & \textbf{ExpectedType} & \textbf{Use Case}  & \textbf{Cardinality} & \textbf{Description}\\ \hline

rai:dataCollection	  &  	sc:Text	  &  	Data life cycle	  &  	ONE	  &  	Description of the data collection process	  \\
\hline
rai:dataCollectionType	  &  	sc:Text	  &  	Data life cycle	  &  	MANY	  &  	"Define the data collection type. Recommended values:
Surveys, Secondary Data analysis, Physical data collection, Direct measurement, Document analysis, Manual Human Curator, Software Collection, Experiments, Web Scraping, Web API, Focus groups, Self-reporting, Customer feedback data, User-generated content data, Passive Data Collection, Others"	  \\
\hline
rai:dataCollectionMissingData	  &  	sc:Text	  &  	Data life cycle	  &  	ONE	  &  	Description of missing data in structured/unstructured form	  \\ 
rai:dataCollectionRawData	  &  	sc:Text	  &  	Data life cycle	  &  	ONE	  &  	Description of the raw data i.e. source of the data	  \\
\hline
rai:dataCollectionTimeframe	  &  	sc:DateTime	  &  	Data life cycle	  &  	MANY	  &  	Timeframe in terms of start and end date of the collection process	  \\
\hline
rai:dataImputationProtocol	  &  	sc:Text	  &  	Compliance	  &  	ONE	  &  	Description of data imputation process if applicable	  \\
\hline
rai:dataManipulationProtocol	  &  	sc:Text	  &  	Compliance	  &  	ONE	  &  	Description of data manipulation process if applicable	  \\
\hline
rai:dataPreprocessingProtocol	  &  	sc:Text	  &  	Data life cycle	  &  	MANY	  &  	Description of the steps that were required to bring collected data to a state that can be processed by an ML model/algorithm, e.g. filtering out incomplete entries etc.	  \\
\hline
rai:dataAnnotationProtocol	  &  	sc:Text	  &  	Data labeling	  &  	ONE	  &  	Description of annotations (labels, ratings) produced, including how these were created or authored - Annotation Workforce Type, Annotation Characteristic(s), Annotation Description(s), Annotation Task(s), Annotation Distribution(s)	  \\
\hline
rai:dataAnnotationPlatform	  &  	sc:Text	  &  	Data labeling	  &  	MANY	  &  	Platform, tool, or library used to collect annotations by human annotators	  \\
\hline
rai:dataAnnotationAnalysis	  &  	sc:Text	  &  	Data labeling	  &  	MANY	  &  	Considerations related to the process of converting the “raw” annotations into the labels that are ultimately packaged in a dataset - Uncertainty or disagreement between annotations on each instance as a signal in the dataset, analysis of systematic disagreements between annotators of different socio demographic group, how the final dataset annotations will relate to individual annotator responses	  \\
\hline
rai:dataReleaseMaintenancePlan	  &  	sc:Text	  &  	Compliance	  &  	MANY	  &  	Versioning information in terms of the updating timeframe, the maintainers, and the deprecation policies.	  \\
\hline
rai:personalSensitiveInformation	  &  	sc:Text	  &  	Compliance	  &  	MANY	  &  	Sensitive Human Attribute(s)- Gender, Socio-economic status,Geography, Language, Age,Culture, Experience or Seniority, Others (please specify)	  \\
\hline
rai:dataSocialImpact	  &  	sc:Text	  &  	AI safety and fairness evaluation	  &  	ONE	  &  	Discussion of social impact, if applicable	  \\
\hline
rai:dataBiases	  &  	sc:Text	  &  	AI safety and fairness evaluation	  &  	MANY	  &  	Description of biases in dataset, if applicable	  \\
\hline
rai:dataLimitations	  &  	sc:Text	  &  	AI safety and fairness evaluation	  &  	MANY	  &  	Known limitations - Data generalization limits (e.g related to data distribution, data quality issues, or data sources) and on-recommended uses.	  \\
\hline
rai:dataUseCases	  &  	sc:Text	  &  	AI safety and fairness evaluation	  &  	MANY	  &  	Dataset Use(s) - Training, Testing, Validation, Development or Production Use, Fine Tuning, Others (please specify), Usage Guidelines. Recommended uses.	  \\
\hline
rai:annotationsPerItem	  &  	sc:Text	  &  	Data labeling	  &  	ONE	  &  	Number of human labels per dataset item	  \\
\hline
rai:annotatorDemographics	  &  	sc:Text	  &  	Data labeling	  &  	MANY	  &  	List of demographics specifications about the annotators	  \\
\hline
rai:machineAnnotationTools	  &  	sc:Text	  &  	Data labeling	  &  	MANY	  &  	List of software used for data annotation ( e.g. concept extraction, NER, and additional characteristics of the tools used for annotation to allow for replication or extension)	  \\
\hline
    \end{tabular}
\end{table*}

  \end{appendices}

\end{document}